\documentclass[twocolumn,noshowpacs,preprintnumbers,amsmath,amssymb]{revtex4-2}
\usepackage{graphicx}
\usepackage{dcolumn}
\usepackage{bm}
\usepackage{epigraph}

\begin{document}

\title{How to write a contemporary scientific article}

\author{Vladimir M. Krasnov}

\affiliation{Department of Physics, Stockholm University, AlbaNova University Center, SE-10691 Stockholm, Sweden}


\begin{abstract}

 Today scientists are drowned in information and have no time for reading all publications even in a specific area. Information is sifted and only a small fraction of articles is being read. Under circumstances, scientific articles have to be properly adjusted to pass through the superficial sifting. Here I present instructions for PhD students with almost serious advises on how to write (and how not to write) a contemporary scientific article. I argue that it should “tell a story” and should answer on the three main questions: Why, What and So what?

\end{abstract}
\maketitle

\epigraph{
\emph{Therefore, since brevity is the soul of wit,}\\
\emph{And tediousness the limbs and outward flourishes,}\\
\emph{I will be brief.}~~~~~~~(W. Shakespeare \cite{Gamlet})
}

\section*{Introduction}
Science has been industrialized. It follows a market-driven Moor's
law \cite{Moore_1965}: the number of scientific publications is
growing exponentially with time
\cite{Larsen_2010,Bornmann_2015,Seppelt_2018}. At the beginning of my
scientific career, slightly more than thirty years ago, we had one ``library day" per week. Then we could skip the lab and go to a library instead. It was crowded in our library. People were sitting there, browsing all newly received journals from the beginning to the end. The
number of such journals could be counted by fingers. Today 24/7
would not be enough for reading all publications
even in my specific area.
I don't even know titles of all relevant journals. Unfortunately,
the dramatic growth of the quantity came at the expense of
the quality \cite{Seppelt_2018}. As a result, the signal-to-noise ratio in scientific literature is reduced. Reading more does not necessarily brings more knowledge. 
I believe that the number of articles read by each researcher did
not increase much in the last three decades. We were reading a lot before and 
have to work as well. To cope with the overflow of information, we use some sifting procedures. Therefore, a contemporary article should be adopted for passing the superficial sifting.

Growing complexity of modern science together with its
industrialization have led to narrowing of research
specializations. We are no longer either experimentalists or
theoreticians, but have a much finer distinction (check e.g.,
academic job announcements). Narrow specialization causes
difficulties in communication between scientists. It is not
uncommon that experts in the same area, sitting in the same
conference room, barely understand each other. Therefore, a
contemporary article should be written in a manner comprehensible 
by not-exactly-the-specialists in {\em your} field, 
which often coincides with the rest of scientific community.

Here I present instructions, that I used to give to my PhD students, on how to
write (and how not to write) a contemporary scientific article. I argue that it should
``tell a story" with a clear and straightforward message and should answer
the three main questions: Why? What? and So what? I hope that these instructions, together with many earlier advises \cite{Popular,Matthews_2014,Menger_2013,Turbek_2016}, can help young scientists in writing more comprehensible papers with better chances to be noted by scientific community.

\section*{Why, What and So What?}

\begin{figure*}[t]
\includegraphics[width=0.99\textwidth]{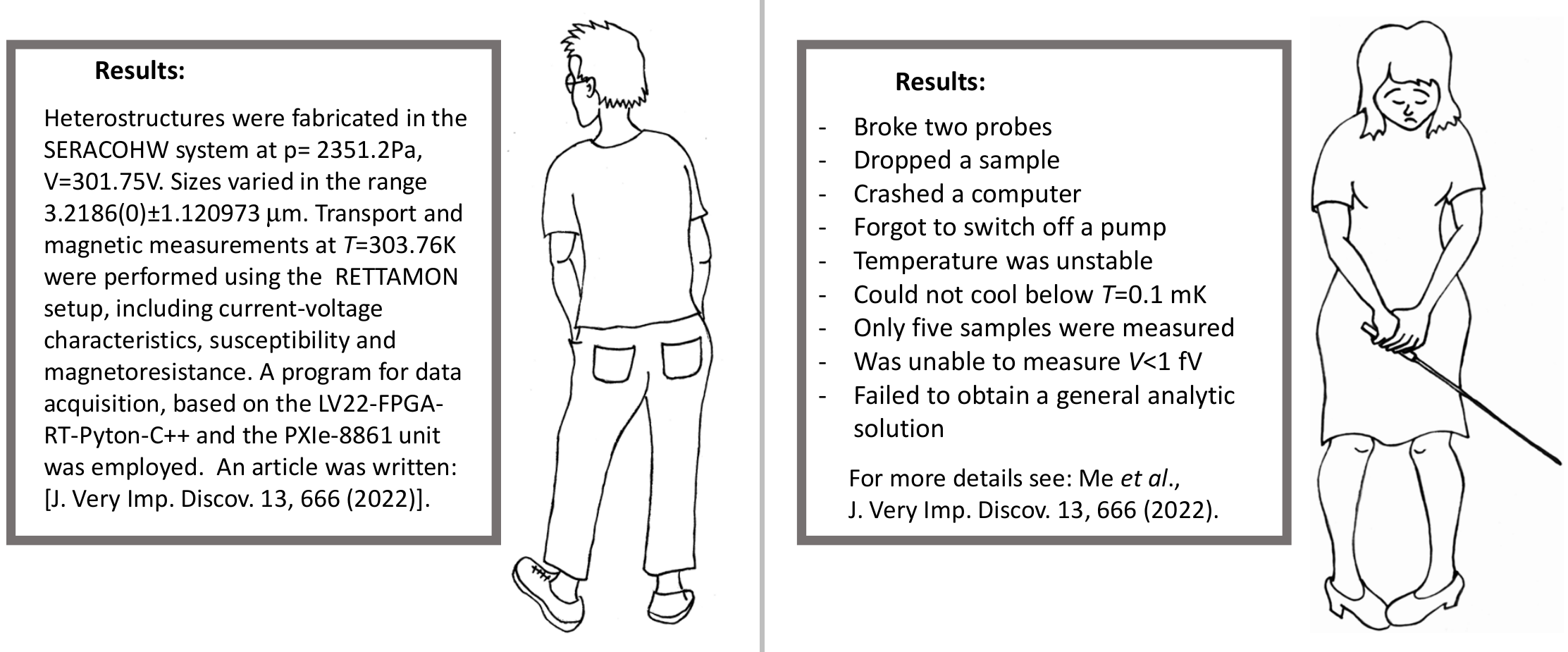}
\label{fig:1} \caption{Typical mistakes made by students: extreme
technicality (left) and extreme negativity (right).}
\end{figure*}

The goal of every author is be read, understood and appreciated. To succeed, first of all, the results should be of the highest scientific value. 
But it is also important that the style is properly adjusted to modern realities.
The article should tell a straightforward and easily digestible story. Like a Hollywood movie, it should contain a prologue, an action and a happy end, which should answer the three main questions posted above. 

Introduction is the prologue of the story. It should explain the
motivation: {\em Why} was it necessary to spend efforts on
this project and why the reader should read it?

Results represent the main action. It should describe {\em What} has been done. The action should not be long and boring
otherwise a spectator will switch it off. ``Brevity is
the soul of wit" \cite{Gamlet}. The story should keep
readers attention. For that, it should have a clear red line,
{\em the message}. 
The action should follow the main story and should not deviate to technical details, or be overloaded with irrelevant data (usually this is the vast majority of acquired experimental results). As in a kitchen: if you put everything in a soup, it will become indigestible. For many students this is counter-intuitive. Technicalities are dear to their hearts because they put so many efforts struggling with them. But it is important to ``see the forest for the trees". The message should be clear without technicalities. They do matter, but only after the paper is read and the message understood. Move them to Appendix, or Supplementary. 

Apart from technicalities, many students tend to focus on problems
and failures. The reason is the same - they represent the most
painful and memorable moments of the project. A report
on a successful work may sound like a complete disaster. The story
must be written in a major key! If there are no successful
results, nor a message to tell then the article should not be
written in the first place. Otherwise, be positive! The reader
doesn't need to know all your mistakes. There are
infinite ways to do things wrong and only one to do it right. Describe how it was done in the end. For example, if the current was too small to be measured directly and an
alternative indirect technique was used for this purpose, don't
write: ``we failed to measure the current". Write instead that ``we
estimated the current from lock-in measurements, as described in Ref.
\cite{Hovhannisyan_2021}". In Figure 1 I sketched the two typical
mistakes.

Finally, Discussion and Conclusions sections represent ``the happy
end". They should answer on the toughest question {\em So What}?
Here the key result, it's novelty and importance have to be explicitly
articulated. However, unlike in a movie, this should
not be the first and only catharsis. For the message to sink in a
human brain it has to be repeated three times.
Therefore, to avoid misunderstanding, there should be: (i) a
spoiler in the Introduction, (ii) a message claim in the Results
and (iii) the moral in Conclusions.

{\bf Title and abstract.}

Today we are not sitting in libraries, but are using internet:
Google, Web of science, e.t.c. This makes the title, the
abstract and the cover art of special importance because they are
passing the first sifting grid. Especially the title. When I was a
student, I was taught to write excruciatingly detailed titles. My
first paper was called: 
``The extended Bean critical state model for superconducting 3-axes ellipsoid and its application for obtaining the bulk critical field $H_{c1}$ and the pinning current $J_c$ in high-$T_c$ superconducting single crystals". 
Informative, isn't it? But today the title should be both informative and
eye-catching. Unfortunately, these two requirements are often
contradictory. Much stretching towards popular catchy titles leaves a bad after-taste. There should be some golden mean. If the choice is between informative and catchy titles, I definitely recommend the informative. Yet, even in
this case there is some flexibility. The title may be informative
e.g. about the key result or the main message (which do not need
to be identical). Keep in mind, that other researchers will be
searching for information on a specific subject. The more closely
your title reflects the content, the more successful will their
search be, increasing the probability of your article
to be read. Google search the chosen title yourself and see if it ends
up in a right category.

Abstract appears at the second step of sifting. It should tell the
story and bring the moral like a fairy tale in one sentence. This
is not easy. The only advice I have - leave abstract writing to
the end, when the first draft is ready and the message is
crystallized.

{\bf Figures.}

At the final stage of sifting, we look through the article (often
from the end) and, just like kids browsing a new fairy tale book,
we focus on Figures. Therefore, Figures should tell a self-consistent story like a comics book.

Students can hardly imagine that in old, pre-computer, times graphs were drawn by hand. Special draftsmen draw axes and symbols. Thanks to computers, modern articles contain much more detailed
visual information. However, a misuse
of computer graphics can lead to crowded and unintelligible
pictures. A general advice: avoid insets and use the minimal
amount of text in the Figures. Imagine that someone would like to
repost a part of your graph. In this case overlapping with
excessive information on the same graph would create a problem.
The modern trend is to have Figures with several simple panels.
This helps to tell a story in a sequential comics-book fashion.

I always recommend to start writing a paper by assembling Figures.
They form a skeleton of a future article, which is then developed
by adding text and description. In experimental work Figures
represent a quintessence of the article. They illustrate results
and carry the message. It is not uncommon that the message is
revealed only after arranging all the Figures.

{\bf References.}

Every scientific journal requires fair representation of earlier
publications, which puts an article in a proper historical
perspective. Therefore, an article should contain a good volume of
references. Too few raze questions if the authors are aware
of the field, is the representation fair, or is the field
important? Excessive self-citations cause irritation.
Self-citations should not exceed 20-25 $\%$ of the total list. Try
to include works from as many different research groups as
possible. Think that the article will be reviewed by several
experts in the field. They wouldn't be happy if their important
(as they think) work is not properly cited. Scientists can be very
petty and picky when it comes to priorities.

The main purpose of a reference is to provide material for deeper
reading on the subject. Make sure that each reference is cited in
a relevant context. Read them all! Topical reviews are the trend of
our time. They are useful for a quick orientation in the field.
Unfortunately they also become a popular lazy reference about
everything. I recommend to be restrictive with reviews. Cite
original articles instead, both for providing focused information
to readers and for giving a scientific respect to pioneers.

{\bf Submission.}

Thoroughly check publication criteria in the chosen journal. Referees are asked to provide answers to specific questions (novelty, originality, impact, ... ). Try to put yourself in the referee's shoes. Count on having at least one referee from outside your field. Things that are obvious to you may not be obvious to the referee. Address the specific questions in the text to help the referee. 

Don't rush with submission. Polish the text very carefully. Don't ignore small details (e.g. mismatch of figure stiles, fonts colors, language, e.t.c.). A good
work written in a sloppy manner will get less credits. You may not have a chance to improve the manuscript afterwards. Let the finished manuscript rest for two weeks. You will likely discover that it reads somewhat differently, the logics is not as
straight as it seemed to be, and the text contains bugs. Repeat
this step until iterations converge and only then press the
submit button.\\

\section*{Conclusions}

I have argued that a modern article should answer the three
main questions. Here I address them to myself:

{\em Why}? Our time with an overflow of information and a narrow
specialization of researchers requires proper adjustment of epistolary
scientific style. Contemporary articles should tell an easily digestible
story with a clear red line and an explicit
message in order to pass the superficial sifting process.

{\em What}? I've wrote down instructions, that I used to give to
my PhD students. By the way, similar rules apply to
conference presentations.

{\em So what}? I hope that presented advises can help students to write more comprehensible articles with a better chance to be noted by scientific community. 
Young scientists should learn the art of clear and laconic expression of
ideas if they want to stay in academia. However, I want to
emphasize that the best strategy for having your paper read is to
maintain a good scientific reputation by not producing ``scientific noise". There are no magic tricks that could make a mediocre research
good. Yet, even a good researcher, presenting excellent
results, should try to help stressed and pressed contemporary
readers.

\end{document}